\documentclass{iaus}
\usepackage{graphicx}
\usepackage{multicol}
\usepackage{amsmath}

\title[Retrograde resonances in compact multi-planetary systems] 
{Retrograde resonances in compact multi-planetary systems:\\
a feasible stabilizing mechanism}

\author[Julie Gayon \& Eric Bois]   
{Julie Gayon
 \and Eric Bois}

\affiliation{Universit\'e Nice Sophia-Antipolis, CNRS, 
  Observatoire de la C\^ote d'Azur,\\ 
  Laboratoire Cassiop\'ee, B.P. 4229, F-06304 Nice Cedex 4, France\\ 
  email: {\tt julie.gayon@oca.eu - 
              eric.bois@oca.eu} \\[\affilskip]
}

\pubyear{2008}
\volume{249}  
\pagerange{1--5}
\jname{Exoplanets: Detection, Formation and Dynamics}
\editors{Y.-S. Sun, S. Ferraz-Mello and J.-L. Zhou, eds.}
\begin{document}

\maketitle

\begin{abstract}
Multi-planet systems detected until now are in most cases characterized by 
hot-Jupiters close to their central star as well as high eccentricities. As a 
consequence, from a dynamical point of view, compact multi-planetary 
systems form a variety of the general $N$-body problem (with $N \ge 3$), 
whose solutions are not necessarily known. Extrasolar planets are up to now 
found in prograde (i.e. direct) orbital motions about their
host star and often in mean-motion resonances (MMR). 
In the present paper, we investigate a theoretical alternative suitable for
the stability of compact multi-planetary systems. When
the outer planet moves on a retrograde orbit in MMR with respect to 
the inner planet, we 
find that the so-called retrograde resonances present fine and
characteristic structures particularly relevant for dynamical stability. We 
show that retrograde resonances and their resources open a family of
stabilizing mechanisms involving specific behaviors of apsidal precessions. We
also point up that for particular orbital data, retrograde MMRs may provide 
more robust stability compared to the corresponding prograde 
MMRs. 

\keywords{celestial mechanics, planetary systems, methods: numerical, statistical}
\end{abstract}

\firstsection 
\section{Introduction}
To identify the dynamical state of multi-planetary systems, we use the MEGNO 
technique (the acronym of Mean Exponential Growth factor of Nearby Orbits; 
Cincotta \& Sim\`o 2000). This method provides relevant information about the 
global dynamics and the fine structure of the phase space, and yields 
simultaneously a good estimate of the Lyapunov Characteristic Numbers with a 
comparatively small computational effort. From the MEGNO technique, we have 
built the MIPS package (acronym of Megno Indicator for Planetary Systems)
specially devoted to the study of planetary systems in their 
multi-dimensional space as well as their conditions of 
dynamical stability.

Particular planetary systems presented in this paper are only used as initial 
condition sources for theoretical studies of 3-body problems. By convention,
the reference system is given by the orbital plane of the inner planet at  
$t = 0$. Thus, we suppose the orbital inclinations and the longitudes of node
of the inner (noted 1) and the outer (noted 2) planets 
(which are non-determined parameters from observations) as follows : $i_1 =
0^\circ$ and $\Omega_1 = 0^\circ$ in such a way that the relative inclination
and the relative longitude of nodes are defined at $t=0$ as follows : 
$i_r = i_2-i_1 = i_2$  and $\Omega_r = \Omega_2-\Omega_1 = \Omega_2$. 
The MIPS maps presented in this paper have been confirmed by 
a second global analysis technique (Marzari {\it et al.} 2006) based on
the Frequency Map Analysis (FMA; Laskar 1993).

\section{Fine structure of retrograde resonance}
\begin{figure}[!h]
\begin{center}
   \includegraphics[width=4.4cm,keepaspectratio=true,angle=270]{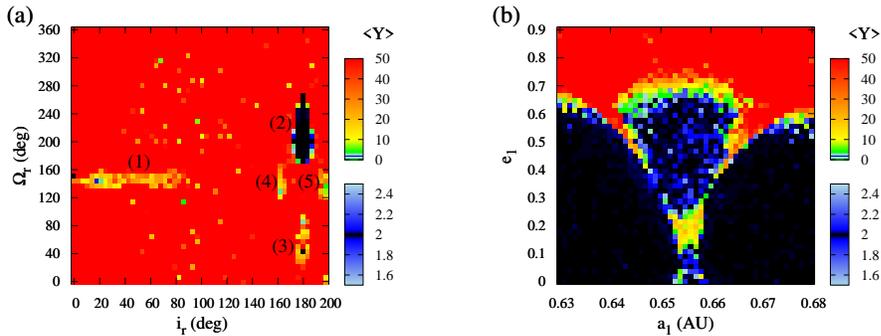}
   \label{fig1}
\end{center}
\caption{Panel (a): Stability map in the $[i_r, \Omega_r]$
  non-determined parameter space of the HD\thinspace73526 planetary system 
  (see Table 1).  
  Panel(b): Stability map in the $[a_1, e_1]$ parameter 
  space for initial conditions taken in the stable zone (2) of panel (a). 
  Note that masses remain untouched whatever the mutual
  inclinations may be; they are equal to their minimal observational values.
  Black and dark-blue colors indicate stable orbits ($<Y>=2 \pm 3\%$ and 
  $<Y>=2 \pm 5\%$ respectively with $<Y>$, the MEGNO indicator 
  value) while warm colors indicate highly unstable orbits.}
\end{figure}

Studying conditions of dynamical stability in the neighborhood of the
HD\thinspace73526 two-planet system (period ratio: 2/1, 
see initial conditions in Table 1), we
only find one stable and robust island (noted (2)) for a 
relative inclination of about $180^\circ$ (see
Fig. \ref{fig1}a). Such a relative inclination (where in fact $i_1=0^\circ$
and $i_2=180^\circ$) may be considered to a coplanar system where the planet
2 has a retrograde motion with respect to the planet 1. From a
kinematic point of view, it amounts to consider a scale change of $180^\circ$ 
in relative inclinations. 
Taking into account initial conditions inside the island (2) 
of Fig. 1a, we show that the presence of a strong mean-motion resonance (MMR)
induces clear stability zones with a nice V-shape structure, as shown
in Fig. 1b plotted in the $[a_1, e_1]$ parameter space.  
Let us note the narrowness of this V-shape, namely only about 0.006 AU wide 
for the inner planet (it is 5 times larger in the Jupiter-Saturn case).
A similar V-shape structure is obtained in $[a_2, e_2]$ with
about 0.015 AU wide. Due to the retrograde motion of the outer planet
2, this MMR is a 2:1 retrograde resonance, also noted 2:-1 MMR.

\begin{figure}[!ht]
\begin{center}
   \includegraphics[width=4.4cm,keepaspectratio=true,angle=270]{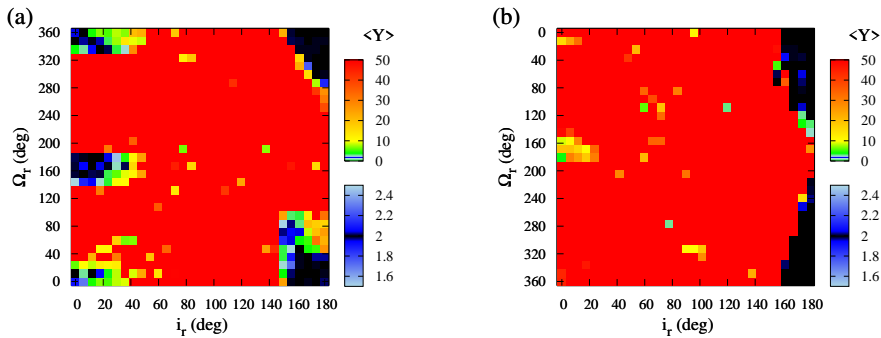}
   \label{fig2}
\end{center}
   \caption{Stability maps in the $[i_r, \Omega_r]$ parameter space. 
     Panel (a): initial HD\thinspace82943 planetary system (see Table 1).
     Panel (b): scale reduction of the HD\thinspace82943 planetary system
     according to a factor 7.5 on semi- major axes. Masses in Panel (a) and
     Panel (b) are identical. Color scale is the same as in Fig. 1.} 
\vspace{3mm}
\end{figure}

\section{Efficiency of retrograde resonances}
Fig. 2 exhibits stability maps in the $[i_r, \Omega_r]$ parameter space
considering a scale reduction of the HD\thinspace82943 planetary system 
(see Table 1) according to a factor 7.5 on semi-major
axes (masses remaining untouched). The dynamical behavior of
the reduced system (Fig. 2b) with respect to the initial one 
(Fig. 2a) points
up the clear robustness of retrograde configurations contrary to prograde 
ones. The ``prograde'' stable islands completely disappear while only the
``retrograde'' stable island  resists, persists and even extends more or less. 
Even for very small semi-major axes and large planetary 
masses,
which should a priori easily make a system
unstable or chaotic, stability is possible with counter-revolving
orbits.  

In the case of the 2:1 retrograde resonance, although close 
approaches happen more often (3 for the 2:-1 MMR) compared to the 2:1 prograde 
resonance, the  2:-1 MMR remains very efficient for
stability because of faster close approaches between the
planets. A more detailed numerical study of retrograde resonances can be found
in Gayon \& Bois (2008).

\begin{table}[!h]
\begin{center}
   \begin{tabular}{|c|ccccc|}
   \hline
   Elements&
   \textrm{HD$\thinspace$73526}&
   \textrm{HD$\thinspace$82943}&
   \textrm{HD$\thinspace$128311}&
   \textrm{HD$\thinspace$160691}&
   \textrm{HD$\thinspace$202206}\tabularnewline
   \hline

    $M_{star}$ $(M_{\odot})$&
    $1.08 \pm 0.05$&
    $1.15$& 
    $0.84$& 
    $1.08 \pm 0.05$&
    $1.15$\tabularnewline
    \hline

   $m \textrm{ sin } i_l \textrm{ ($M_J$)}$&
   $\begin{array}{c}
   2.9 \pm 0.2\\
   2.5 \pm 0.3
   \end{array}$&
   $\begin{array}{c}
   1.85 \\
   1.84 
   \end{array}$&
   $\begin{array}{c}
   1.56 \pm 0.16\\
   3.08 \pm 0.11 
   \end{array}$&
   $\begin{array}{c}
   1.67 \pm 0.11\\
   3.10 \pm 0.71
   \end{array}$&
   $\begin{array}{c}
   17.4 \\
   2.44
   \end{array}$\tabularnewline
   \hline

   $a \textrm{ (AU)}$&
   $\begin{array}{c}
   0.66 \pm 0.01\\
   1.05 \pm 0.02
   \end{array}$&
   $\begin{array}{c}
   0.75 \\
   1.18
   \end{array}$&
   $\begin{array}{c}
   1.109 \pm 0.008 \\
   1.735 \pm 0.014 
   \end{array}$&
   $\begin{array}{c}
   1.50 \pm 0.02\\
   4.17 \pm 0.07
   \end{array}$&
   $\begin{array}{c}
   0.83 \\
   2.55
   \end{array}$\tabularnewline
   \hline

   $e$&
   $\begin{array}{c}
   0.19 \pm 0.05\\
   0.14 \pm 0.09
   \end{array}$&
   $\begin{array}{c}
   0.38 \pm 0.01\\
   0.18 \pm 0.04
   \end{array}$&
   $\begin{array}{c}
   0.38 \pm 0.08 \\
   0.21 \pm 0.21 
   \end{array}$&
   $\begin{array}{c}
   0.20 \pm 0.03\\
   0.57 \pm 0.1
   \end{array}$&
   $\begin{array}{c}
   0.435 \pm 0.001\\
   0.267 \pm 0.021 
   \end{array}$\tabularnewline
   \hline

   $\omega \textrm{ (deg)}$&
   $\begin{array}{c}
   203 \pm 9\\
   13 \pm 76
   \end{array}$&
   $\begin{array}{c}
   124.0 \pm 3\\
   237.0 \pm 13
   \end{array}$&
   $\begin{array}{c}
   80.1 \pm 16\\
   21.6 \pm 61 
   \end{array}$&
   $\begin{array}{c}
   294 \pm 9\\
   161 \pm 8
   \end{array}$&
   $\begin{array}{c}
   161.18 \pm 0.30 \\
   78.99 \pm 6.65 
   \end{array}$\tabularnewline
   \hline

   $M \textrm{ (deg)}$&
   $\begin{array}{c}
   86 \pm 13\\
   82 \pm 27
   \end{array}$&
   $\begin{array}{c}
   0\\
   75.21 \pm 1.96
   \end{array}$&
   $\begin{array}{c}
   257.6 \pm 2.7\\
   166 \pm 2
   \end{array}$&
   $\begin{array}{c}
   0 \\
   12.6 \pm 11.2
   \end{array}$&
   $\begin{array}{c}
   105.05 \pm 0.48 \\
   311.6 \pm 9.5 
   \end{array}$
   \tabularnewline
   \hline

\end{tabular}
\begin{tabular}{ll}
\end{tabular}
   \caption{\label{tab1}Orbital parameters of the HD$\thinspace$73526,
     HD$\thinspace$82943, HD$\thinspace$128311, HD$\thinspace$160691 and
     HD$\thinspace$202206 planetary systems. Data sources come from Tinney et
    al. (2006), Mayor et al. (2004), Vogt et al. (2005), McCarthy et
    al. (2004) and Correia et al. (2005) respectively. For each system and
    each orbital element, the first line corresponds to the inner planet and
    the second one to the outer planet.}

\end{center}
\end{table}

\begin{table}[!h]
\begin{center}
   \begin{tabular}{cccc}\hline
   \hspace{0.2cm}Data sources\hspace{0.2cm} & 
   \hspace{0.2cm}Period ratio\hspace{0.2cm} &   
   \hspace{0.2cm}Prograde MMR\hspace{0.2cm} &  
   \hspace{0.2cm}Retrograde MMR\hspace{0.2cm}\\ \hline 
   HD\thinspace73526 & 2/1&    17 &   500 \\
   HD\thinspace82943 & 2/1&   755 &  1000 \\
   HD\thinspace128311& 2/1&   249 & 137 \\
   HD\thinspace160691& 5/1& $\varepsilon$ &   320 \\
   HD\thinspace202206& 5/1& $\varepsilon$ &  631 \\ \hline
\end{tabular}
 \label{tab2}
  \caption{Statistical results. For each type of MMR (prograde or retrograde), 
    1000 random systems have been integrated in the error bars of each data
    source. The proportion of stable systems over 1000 is indicated in each 
    case. $\varepsilon$ designates a very small value that depends on the 
    size of the random system size. Data sources come from  
    Tinney et al. (2006), Mayor et al. (2004), Vogt et al. (2005), McCarthy et
    al. (2004) and Correia et al. (2005) respectively (see Table 1).}
\end{center}
\end{table}

\section{Occurrence of stable counter-revolving configurations}
The occurence of stable two-planet systems including
counter-revolving orbits  appears in the neighborhood of a
few systems observed in 2:1 or 5:1 MMR. 
New observations frequently induce new determinations of 
orbital elements. It is the case for the HD\thinspace160691 planetary system
given with 2 planets in McCarthy {\it et al.} (2004) then with 4 planets in
Pepe {\it et al.} (2007). Hence, systems related to initial conditions 
used here (see Table 1) 
have to be considered as {\it academic} systems. Statistical results for
stability of these academic systems are presented in Table 2, both
in the prograde case ($i_r=0^\circ$) and in the retrograde case 
($i_r=180^\circ$). For each data source, 1000 random 
systems taken inside observational error bars have been 
integrated. Among these random systems, the proportion of stable systems
either with prograde orbits or with counter-revolving orbits is
given in Table 2. In all cases, a significant number of stable systems is 
found in retrograde MMR. Moreover, in most data sources, 
retrograde possibilities predominate.

\section{Resources of retrograde resonances}
The 2:1 (prograde) MMRs preserved by synchronous precessions of the apsidal 
lines (ASPs) are from now on well understood
(see for instance Lee \& Peale 2002, Bois {\it et al.} 2003,
  Ji {\it et al.} 2003, Ferraz-Mello {\it et al.} 2005). The MMR-ASP 
combination is often very effective; however, ASPs may also exist alone for
stability of planetary systems. 
Related to subtle relations between the eccentricity of the inner orbit 
($e_1$) and the relative apsidal longitude $\Delta\tilde{\omega}$ 
(i.e. $\tilde{\omega}_1-\tilde{\omega}_2$), Fig. 3 
permits to observe how the 2:1 retrograde MMR brings out 
its resources in the
$[\Delta{\tilde{\omega}}, e_1]$ parameter space :

\begin{itemize}
\item In the island (1) (i.e. inside the $[a, e]$ V-shape of Fig. 1b), the 
2:-1 MMR is combined with a uniformly prograde ASP (both planets precess on
average at the {\it same rate} and in the {\it same prograde direction}).

\item In the island (2) (i.e. outside but close to the $[a, e]$ V-shape of 
  Fig. 1b), the 2:-1 {\it near}-MMR is combined with a particular apsidal
  behavior that we have called a {\it rocking} ASP (see Gayon \& Bois 2008):
  both planets precess at the {\it same rate} but in {\it opposite directions}.

\item The $[\Delta\tilde{\omega}, e_1]$ map also exposes a
third island (3) that proves to be a wholly chaotic zone on
long term integrations.
\end{itemize}
Let us note that the division between islands (1) and (2) is related to the 
degree of closeness to the 2:-1 MMR. 

\begin{figure}[h]
\begin{center}
\begin{multicols}{2}
   \includegraphics[width=4.4cm,keepaspectratio=true,angle=270]{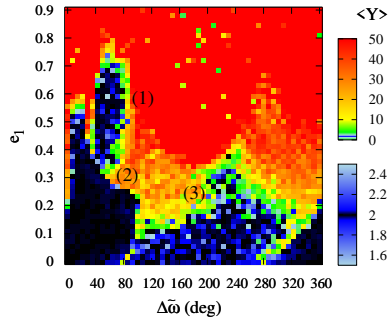}
\columnbreak
\\
$\!$
\vspace{1.4cm} 
\caption{Stability map in the $[\Delta{\tilde{\omega}}, e_1]$ parameter
  space. A similar distribution of stable islands is obtained in
  $[\Delta{\tilde{\omega}}, e_2]$. Color scale and initial
    conditions are the same as in Fig. 1 with in addition the
      $i_r$ and $\Omega_r$ values chosen in the island (2) of Fig. 1a.} 
\end{multicols}
   \label{fig3}
\end{center}
\end{figure}

\section{Conclusion}
We have found that retrograde resonances present fine and characteristic
structures particularly relevant for dynamical stability. We have also shown 
that in cases of very compact systems obtained by scale reduction, only the
"retrograde" stable islands survive. From our statistical
approach and the scale reduction experiment, 
we have expressed the efficiency for stability of
retrograde resonances. Such an efficiency can be understood  
by very fast close approaches between the planets 
although they are in greater number. 

We plan to present an Hamiltonian approach of retrograde MMRs 
in a forthcoming paper (Gayon, Bois, \& Scholl, 2008). Besides, in Gayon 
\& Bois (2008), we propose
two mechanisms of formation for systems harboring counter-revolving orbits. 
Free-floating planets or the Slingshot model might indeed explain
  the origin of such planetary systems.

In the end, we may conclude that retrograde resonances prove to be a feasible stabilizing 
mechanism.

\acknowledgements{We thank the anonymous referee for his 
  comments that greatly helped to improve the paper.}

\end{document}